\documentclass{style}
\copyrightyear{2014}
\pubyear{2014}
\usepackage[usenames,dvipsnames,svgnames,table]{xcolor}

\begin{document}
\firstpage{1}

\title[Integrating alignment-based and alignment-free sequence similarity measures]{Integrating alignment-based and alignment-free sequence similarity measures for biological sequence classification}
\author[Borozan \textit{et~al}]{Ivan Borozan\footnote{to whom correspondence should be addressed}, Stuart Watt and Vincent Ferretti}
\address{Ontario Institute for Cancer Research, MaRS Centre, South Tower, 101 College Street, Suite 800, Toronto, Ontario, Canada.}
\history{Received on June 5, 2014; revised on January 4, 2015; accepted on January 5, 2015}
\editor{Associate Editor: John Hancock}
\maketitle
\begin{abstract}
\section{Motivation:}
Alignment-based sequence similarity searches, while accurate for some type of sequences, can produce incorrect results when used on more divergent but functionally related sequences that have undergone the sequence rearrangements observed in many bacterial and viral genomes. Here, we propose a classification model that exploits the complementary nature of alignment-based and alignment-free similarity measures with the aim to improve the accuracy with which DNA and protein sequences are characterized. 
\section{Results:}
Our model classifies sequences using a combined sequence similarity score calculated by adaptively weighting the contribution of different sequence similarity measures. Weights are determined independently for each sequence in the test set and reflect the discriminatory ability of individual similarity measures in the training set. Since the similarity between some sequences is determined more accurately with one type of measure rather than another, our classifier allows different sets of weights to be associated with different sequences. Using five different similarity measures we show that our model significantly improves the classification accuracy over the current composition and alignment based models, when predicting the taxonomic lineage for both short viral sequence fragments and complete viral sequences. We also show that our model can be used effectively for the classification of reads from a real metagenome dataset as well as protein sequences. 
\section{Availability and Implementation:}
\begin{sloppypar}
\noindent 
All the datasets and the code used in this study are freely available at https://collaborators.oicr.on.ca/vferretti/borozan\_csss/csss.html.
\end{sloppypar}
\section{Contact:} \href{ivan.borozan@gmail.com}{ivan.borozan@gmail.com}
\section{Supplementary Information:} Supplementary data are available at Bioinformatics online. 

\end{abstract}

\section{Introduction}
Sequence comparison of genetic material between known and unknown organisms plays a crucial role in metagenomic and phylogenetic analysis. Sequence similarity search is a method of sequence analysis that is extensively used for characterizing unannotated sequences (\citealp{Altschul1997}). It consists of aligning a query sequence to a sequence database with the aim of determining those sequences that have statistically significant matches to that of the query. In this way, for example, a known biological function or taxonomic category of the closest match can be assigned to the query for its characterization. 

Alignment-based methods however can produce incorrect results when applied to more divergent but functionally related sequences that have undergone sequence rearrangements. Sequence rearrangements such as genetic recombination and shuffling or horizontal gene transfer are observed in a variety of organisms including viruses and bacteria (\citealp{Delviks-Frankenberry2011,Shackelton2004,Domazet-Loso2011}). These processes, which produce alternating blocks of sequence material, are at odds with the alignment-based sequence comparison which assumes conservation of contiguity between homologous segments (\citealp{Vinga2003}). Another weakness of the alignment-based approach is in the use of different methods for scoring pairwise protein sequence alignments, as reported in \citealp{Vinga2003}.

In addition to sequence rearrangements, viral genomes exhibit gene gain and loss, gene duplication and high sequence mutation rates (\citealp{Shackelton2004,Duffy2008}). The cumulative effect of these changes make viral genomes among the most variable in nature. Because of this high sequence divergence and the often small number of genes, viral genomes present a greater challenge to phylogenetic classification and taxonomic analysis when these are based on sequence comparison by alignment only. Improving the results of such studies is important for better understanding viruses and their involvement in human diseases, including cancer (\citealp{zurHausen2007}). 

Because of these shortcomings active research has been conducted into alignment-free measures to overcome the above limitations. A number of alignment-free measures have been proposed in recent years as reported in two comprehensive reviews (\citealp{Vinga2003,Vinga2014}). 

In this study we propose a new classification model that combines similarity scores obtained from alignment-free and alignment-based similarity measures with the aim to exploit the complementary nature of these measures to improve the classification accuracy. In our model the classification of sequences is performed by using a combined sequence similarity score that is calculated based on the weighted contribution of similarity scores, where weights reflect the discriminatory ability of individual measures in the training set. One unique feature of our model is based on the observation that the similarity between some sequences is determined more accurately with one type of similarity measure rather than another, hence in our model different sets of weights can be associated with different sequences (i.e. sequences to be classified). Furthermore we provide a mathematical framework that can include any number of additional similarity measures and show that our model  {\it{i)}} is applicable to both nucleotide and amino acid sequences {\it{ii)}} improves the classification accuracy over a purely alignment-based sequence comparison approach and {\it{iii)}} improves the classification accuracy for metagenomic analysis of short reads produced  by next generation sequencing technologies.

Recently a number of methods for metagenomic analysis have been proposed (\citealp{Huson2007,Huson2014,Rosen2011,Patil2011,Wood2014,Brady2009,Nalbantoglu2011}). Of these seven methods, PhymmBL (\citealp{Brady2009}) is the method closest in approach to the method presented in this study, since it classifies reads (or contigs) using an integrated score obtained by combining the Interpolated Markov Models score (IMM) (an alignment-free/composition-based similarity measure) with the BLAST score. PhymmBL (\citealp{Brady2009}) has been shown to outperform MEGAN (\citealp{Huson2007}) for longer contigs while for shorter ones the results of comparison are misleading at best since MEGAN produces results in a form that can not be directly compared to those of PhymmBL (\citealp{Brady2009,Brady2011}) and the model proposed in this study. 
We believe that improving the classification accuracy for shorter reads (100bp-1000bp) is critical, since such metagenomic analysis does not require the assembly of raw sequenced reads prior to classification. For these reasons and to address the objective {\it{iii)}} in the previous paragraph, we chose to compare the classification results obtained with the model presented in this study to four primarily composition-based models (PhymmBL (\citealp{Brady2009}), NBC (\citealp{Rosen2011}), PhyloPythiaS (\citealp{Patil2011}) and RAIphy (\citealp{Nalbantoglu2011})) and the two most recently published methods for the classification of metagenomic sequences, Kraken (\citealp{Wood2014}) based on the exact alignment of k-mers, and PAUDA (\citealp{Huson2014}) an alignment based method.

\begin{methods}

\section{Methods}

\subsection{Sequence similarity measures}
In this section we describe the five sequence similarity measures that we chose to use in our classification model. Three of them are alignment free sequence similarity measures and two of them are alignment-based sequence similarity measures.

\subsubsection{Alignment free sequence similarity measures:}
The choice of the three alignment-free sequence similarity measures (see below) is based on the notion of complementarity between these measures and the two alignment-based similarity measures that we chose to use in this study. Specifically, similarity measures based on k-mer frequencies (the Euclidean Distance and Jensen-Shannon Divergence) do not depend on any assumption of the contiguity of conserved segments, as the alignment-based measures do. They do, however, depend on the choice of the k-mer size (\citealp{Wu2005}). In contrast, the Compression Based measure (\citealp{Li2001}) built upon the concept of Kolmogorov complexity is both independent of the k-mer size (since it is not based on k-mer counts) and the assumption of the contiguity of conserved segments. 

The Euclidean Distance and Jensen-Shannon Divergence measures both require the number of all possible k-mers ${K= n^k}$ to be counted for any given sequence, where $n$ is the alphabet size (i.e. $n$ = 4 for DNA sequences and $n$ = 20 for protein sequences) and $k$ is the length of the k-mer sequence. To count the number of k-mers in DNA sequences we use the JELLYFISH (\citealp{Marcais2011}) algorithm and for protein sequences we use a Python script from \citealp{Gupta2008}. The raw counts are used to form a vector ${\bf{C^k}}$ of all possible k-mers of length $k$,

\begin{equation}
{\bf{C^k}} = <c^k_1, c^k_2, ..., c^k_K>      		
\end{equation}

raw counts in eq.1 are then normalized to form a probability distribution vector 

\begin{equation}
{\bf{F^k}} = {\bf{C^k}} /\sum\limits_{i=1}^K c^k_i = <f^k_1,f^k_2,...,f^k_K>
\end{equation}

giving the relative abundance of each k-mer.

\noindent  
1. {\it{The Euclidean Distance (ED)}}: The similarity score between two sequences $X$ and $Y$ is the Euclidean distance between their $n^k$ dimensional probability distribution vectors  ${\bf{F^k_X}}$ and ${\bf{F^k_Y}}$ as defined in eq.3

\begin{equation}				
d_{ED} = \sqrt{N({\bf{F^k_X}},{\bf{F^k_X}}) - 2N({\bf{F^k_X}},{\bf{F^k_Y}}) + N({\bf{F^k_Y}},{\bf{F^k_Y}})}
\end{equation}

\begin{equation}
N({\bf{X}},{\bf{Y}}) = \frac{{\bf{X}} \cdot {\bf{Y}}}{\sqrt{({\bf{X}} \cdot {\bf{X}})({\bf{Y}} \cdot {\bf{Y}})}}  
\end{equation}

\noindent  
where eq.4 ensures that each vector is normalized and has length 1 in the $n^k$ dimensional space. The choice for this metric is based on its simplicity, well defined mathematical properties and its demonstrated effectiveness as an alternative to the alignment method (\citealp{Vinga2003}). The Euclidean Distance defined in eq.3 has values that range between zero and one, with lower values indicating increasing similarity and higher values decreasing similarity.  

\noindent  
2. {\it{The Jensen-Shannon Divergence (JSD)}}: This is an information theoretic non-symmetric divergence measure of two probability distributions. The Jensen-Shannon divergence between two sequences $X$ and $Y$ is calculated between their $n^k$ dimensional probability distribution vectors ${\bf{F^k_X}}$ and ${\bf{F^k_Y}}$ as shown below 

\begin{equation}	 
d_{JSD} = JSD({\bf{F^k_X}}, {\bf{F^k_Y}}) = 0.5 \cdot KL({\bf{F^k_X}}, {\bf{M}}) +  0.5 \cdot KL({\bf{F^k_Y}}, {\bf{M}})  	
\end{equation}

\noindent  
where M$_i$ = ${(f^k_{x_i} + f^k_{y_i})/2}$, and where $i$ = ${1,...,K}$, and where $KL$ is the Kullback-Leibler divergence defined below

\begin{equation}	 
KL ({\bf{F^k_X}}, {\bf{M}}) =\sum\limits_{i=1}^K f^k_{x_i} \cdot log(f^k_{x_i} / M_i)
\end{equation}

\noindent  
Provided that the base 2 logarithm is used in eq.6, JSD has values that range between zero and one, with lower values indicating increasing similarity and higher values decreasing similarity. The choice for this similarity measure is based on its ability to successfully reconstruct phylogenies using whole-genome sequences as reported in \citealp{Sims2009}. 

\noindent  
3. {\it{The Compression Based (CB) measure}}: This similarity measure is based on the concept of Kolmogorov complexity. Conditional Kolmogorov complexity ${K(X|Y)}$ (or algorithmic entropy) of sequence $X$ given sequence $Y$ is defined as the length of the shortest program computing $X$ on input $Y$. In this way ${K(X|Y)}$ measures the randomness of $X$ given $Y$. The Kolmogorov complexity $K(X)$ of a sequence $X$ is defined as ${K(X|e)}$ where $e$ is an empty string. We note that Kolmogorov complexity $K(X)$ of a sequence $X$ is non-computable and that in practice $K(X)$ is  approximated by the length of the compressed sequence $X$, obtained using compression algorithms such as LZMA or GenCompress (\citealp{Chen1999}). Our choice for this measure is based on the following two properties {\it{i)}} CB is not affected by sequence rearrangements and {\it{ii)}} since CB is not a frequency based measure, it is not affected by the choice of the k-mer size. To calculate the compression-based distance between two sequences $X$ and $Y$ we chose to use the normalized compression distance (NCD) (\citealp{cilibrasi2005clustering}) as defined below:

\begin{equation}
d_{CB} = 1 - NCD(X, Y)
\end{equation}

where

\begin{equation}
NCD(X, Y) = \frac{C(XY) - min\{C(X),C(Y)\}}{max\{C(X), C(Y)\}}
\end{equation}

\noindent  
where $C(.)$ denotes the length of a compressed sequence using a particular compression algorithm and where $XY$ denotes the concatenation of sequence $X$ with sequence $Y$. Note that the NCD in eq.8 is an empirical approximation of the normalized information distance (NID) which is defined as a metric in \citealp{cilibrasi2005clustering}. The distance calculated using eq.7 takes values between zero and one, with lower values indicating increasing sequence similarity and higher values decreasing sequence similarity. The compression algorithm used in this study is plzip (http://www.nongnu.org/lzip/plzip.html) a multi-threaded, lossless data compressor based on the lzlib compression library that implements a simplified version of the LZMA algorithm. All sequences in this study were compressed using plzip with the compression level parameter set to 4, with matched length set to 3MB and dictionary size set to 12 bytes.

\subsubsection{Alignment-based sequence similarity measures:} 
\paragraph{}
4. {\it{The BLAST based measure}}: For the classification of DNA sequences the distance between the query sequence $X$ and subject $Y$ is expressed in terms of the BLAST bit scores calculated using the BLAST algorithm (\citealp{Altschul1997}) (blastall version 2.2.18, blastall -p blastn), with default parameter value settings.\\
\noindent  
5. {\it{The Smith-Waterman based measure}}: For the classification of protein sequences, similarity scores expressed in terms of p-values calculated using the Smith-Waterman algorithm, were taken from \citealp{Liao2003}.  

\subsection{Classification model}
As mentioned in the introduction, we propose to exploit the complementary properties of the 5 individual similarity measures described above in order to improve the accuracy with which nucleotide or amino acid sequences are characterized. Our aim is to propose a combined sequence similarity score (CSSS) that will improve upon the limitations of the individual sequence similarity scores (as described in the introduction section) and lead to an improved classification performance.

The CSSS model rests on three assumptions {\it{i)}} that similarity measures are complementary in nature (as described in the previous section), {\it{ii)}} that some sequences are better characterized with one type of similarity measure than another, and {\it{iii)}} that their individual values are in the range between zero and one.

Among many machine learning algorithms that are available today, the nearest neighbour algorithm (NN) is one of the simplest and most intuitive classification algorithms. For this reason the nearest neighbour algorithm is often used as the reference classifier in comparative studies. The k-nearest neighbours (k-NN) algorithm performs the classification by identifying the $k$ nearest neighbours that are the closest in terms of a distance/similarity measure to a query (or test sample). It then assigns to the query the class that occurs the most often among the $k$ nearest neighbours. In the case where $k$ = 1 the query is assigned the class of the closest single nearest neighbour. Because of these properties we find the 1-NN algorithm to be a natural choice for the classifier in our approach, as described below. 

Let ${\bf{S^j_X}}$ = $<$ $s^j_{x1}$, $s^j_{x2}$, ... , $s^j_{xn}$ $>$  be an $n$ dimensional vector of sequence similarities/distance scores $s^j_{xi}$ between the sequence $X$ in the test set and the $i$th sequence in the training set, calculated using $j$th sequence similarity measure. 

\noindent  
For each sequence $X$ in the test set we can now define the $n$ dimensional ${\bf{S^c_X}}$ vector of combined sequence similarity/distance scores, to be the linear combination of ${\bf{S^j_X}}$ vectors across j = ${\{1,...,J\}}$ similarity measures as shown below

\begin{equation}	 	
{\bf{S^c_X}} = \frac{\sum\limits_{j=1}^{J} w_j \cdot {\bf{S^j_X}}}{\sum\limits_{j=1}^{J} w_j} 
\end{equation}

\noindent  
where $w_j$ is the weight of the $j$th sequence similarity measure calculated as the ratio of the between group variability ($\hat{S}^2_B$) to the within group variability ($\hat{S}^2_W$) (i.e. the F-test statistics) for each ${\bf{S^j_X}}$ vector as shown in eq.10.  

\begin{equation}	 
w_j = \frac{\hat{S}^2_B({\bf{S^j_X}})}{\hat{S}^2_W({\bf{S^j_X}})}
\end{equation}

\noindent  
Note that the combination of scores obtained using different similarity measures shown in eq.9 is performed independently for each sequence $X$ in the test set.

\noindent  
The between group variability $\hat{S}^2_B$ in eq.10 is defined as

\begin{equation}	
\hat{S}^2_B({\bf{S^j_X}}) = \frac{\sum\limits_{cl=1}^{CL}n_{cl}(\overline{S^j_{X_{cl\cdot}}} - \overline{S^j_{X}})^2}{(CL - 1)}  
\end{equation}

\noindent 
where $CL$ denotes the total number of classes (or groups) in the training set, ${\overline{S^j_{X_{cl\cdot}}}}$ denotes the mean of similarity/distance scores in the $cl$th class for the measure $j$, and $n_{cl}$ is the number of observations (or similarity/distance scores) in the $cl$th class.

\noindent 
The within group variability $\hat{S}^2_W$ in eq.10 is defined as 

\begin{equation}
\hat{S}^2_W({\bf{S^j_X}}) = \frac{\sum\limits_{cl,l}(S^j_{X_{cl,l}} - \overline{S^j_{X_{cl\cdot}}})^2}{(N - CL)} 
\end{equation}

\noindent 
where $S^j_{X_{cl,l}}$ is the $l$th similarity/distance score in the $cl$th out of $CL$ classes of $S^j_X$ for the measure $j$, and $N$ is the total number of sequences (or samples) in the training set.

Thus if $X$ represents an unknown sequence in the test set, the k-NN algorithm will find the $k$ nearest examples in the $n$ dimensional vector ${\bf{S^c_X}}$ = $<$ $s^c_{x1}$, $s^c_{x2}$, ... , $s^c_{xn}$ $>$ , where $n$ is the total number of examples with known labels in the training set and $s^c_{xi}$ is the combined similarity/distance score between the sequence $X$ in the test set and the $i$th sequence in the training set.

Prior to combining alignment-based scores (AB) (such as the ones obtained with Smith-Waterman (SW) or BLAST algorithms) with those obtained using alignment-free similarity measures, the $n$ dimensional vector of sequence similarities/distance scores ${\bf{S^{AB}_X}}$ = $<$ $s^{AB}_{x1}$, $s^{AB}_{x2}$, ... , $s^{AB}_{xn}$ $>$, is first transformed into normalized scores as shown in eqs 13 and 14 so that their values range between zero and one, 

\begin{equation}	   
{\bf{S^{BLAST}_X\_norm}} = 1 - \frac{{\bf{S^{BLAST}_X}}}{max\{{\bf{S^{BLAST}_X}}\}}
\end{equation}

\begin{equation}	   
{\bf{S^{SW}_X\_norm}} = \frac{{\bf{S^{SW}_X}}}{max\{{\bf{S^{SW}_X}}\}}
\end{equation}

with lower values indicating increasing sequence similarity and higher values decreasing sequence similarity.

In Figure.1 we illustrate how the combination of sequence similarity scores proposed in this study and a 1-NN classifier can improve the classification accuracy of a given test sample. Let M1 and M2 be two similarity measures, and T a test sample that can be assigned either one of the two classes in the training set (either a `circle' or a `triangle') based on a single nearest neighbour closest in distance to T. Let also assume that T is known to belong to the class `circle'. As shown in Figure1.a, according to the M1 measure the test sample T is assigned the correct class (i.e. `circle') while according to the M2 measure T is assigned the incorrect class (i.e. a `triangle'). In Figure1.b we show how by doing a simple arithmetic mean of distances/scores (i.e. M1 and M2 have same weights) the bias of M2 can be corrected by the M1 measure. Moreover in Figure 1.c we show how a properly weighted arithmetic mean (in this example M1 was assigned a weight of 10 and M2 a weight of 2) can even further improve the classification accuracy of T. We also see from this simple example that 1-NN is the simplest and the most intuitive choice for the classifier for our model since its assigns T to the class based on the single nearest (in terms of a distance) neighbour in the training set. 

\begin{figure}[!tbp]
\centerline{\includegraphics[width=3.0in]{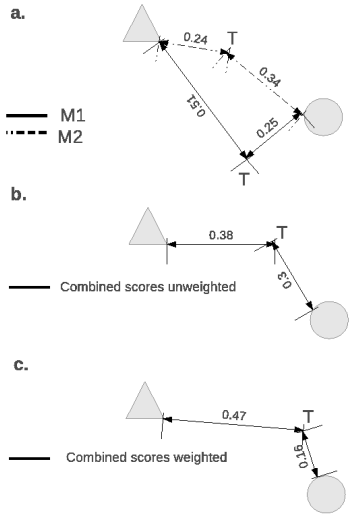}}
\caption{
Shows a graphical representation of how the CSSS model can improve the classification accuracy of a given test sample. M1 and M2 are two similarity measures, and T is a test sample that can be assigned either a class `circle' or a class `triangle' that are present in the training set. In Figure1.a according to M1, T is assigned the correct class (i.e. `circle') while according to M2, T is assigned the incorrect class (i.e. a `triangle'). In Figure1.b we show the classification according to the the combined unweighted score. In Figure 1.c we show the classification according to the combined weighted score. 
}\label{fig:01}
\end{figure}

In the Results section we demonstrate how the choice of different measures (see the previous section) and the weighting scheme proposed in eq.10 leads to an improved classification accuracy on the three different datasets used in this study.

\subsubsection{Selection of similarity measures prior to classification:}
In order to remove similarity measures from eq.9 with low predictive power (for more details see the Discussion section) the selection of similarity measure prior to classification of test samples is performed as follows:

\begin{itemize}
\item the training set is split into two sets, set A and set B, using 2/3, 1/3 splits
\item the classification performance of each similarity measure is evaluated on the set B using set A as the training set
\item only similarity measures with prediction accuracy greater or equal to 10\% are selected 
\end{itemize}

\subsubsection{Selection of the k-mer size:}
The k-mer size necessary to calculate similarity measures in eq.3 and eq.5 is a free parameter in our model and its upper bound needs to satisfy the inequality given in eq.15

\begin{equation}	   
n{^{k}} < L  
\end{equation}

where $n$ is the alphabet size and $L$ is the length of the smallest genome in either the training or test sets. 
The inequality in eq.15 avoids calculations with k-mer sizes that are so large they produce erroneous and artificial differences between genomes that ultimately correlate with genome lengths rather than genome content as described in \citealp{Akhter2013}. Thus to compare genomes (or protein sequences) based on similarity measures (see eqs 3 and 5) that use frequency distribution vectors (see eq.2) the k-mer size should be chosen in such a way to satisfy the inequality shown in eq.15.

\subsection{Datasets}
We evaluate the ability of our model to classify different types of biological sequences using three datasets, one containing viral nucleotide sequences, a second consisting of longueur nucleotide reads (with an average of 759 bp in length) from a real metagenome, and a third consisting of protein sequences.

\subsubsection{Dataset I.}
Due to their considerable variability viral genomes are expected to pose a greater challenge to phylogenetic classification than genomes from other organisms. In this regard we evaluate the classification performance of the CSSS model using a dataset composed of 1066 complete viral genomes downloaded from the NCBI RefSeq database. The classification of viral genomes into genera was performed in three steps:
\noindent  
\newline 
Step 1: the 1066 viral sequences across 147 different genera were divided into training and test sets in such a way that for each genus the test set consisted of viral genomes that were not represented in the training set. The relative sizes of the training and test sets were respectively set to 3/4 and 1/4. 
\newline 
\noindent  
Step 2: selection of similarity measures prior to the classification of test examples selected in step 1 was performed as described in the Methods section (see Selection of similarity measures prior to classification) using the training set from step 1 
\noindent  
\newline 
Step 3: Prediction of viral genera in the test set selected in step 1 is performed using the training set selected in step 1 and combined sequence similarity scores calculated using the formula shown in eq.9 with similarity measures selected in step 2. Note that the training set in this step consists only of complete viral genomes. 

The complete evaluation of the classification performance was carried out using test samples generated in step 1 composed of complete viral genomes and viral sequence fragments of 1000 bp, 500 bp and 100 bp in length, viral fragments were sampled at random from each complete viral genome in the test set obtained in step 1. Note that for viral fragments the set B in step 2 (see also Selection of similarity measures prior to classification in Methods section) contains sequences that are of the same fragment length as those in the test set of step 1. 
 To evaluate the variability of results, training and test sets were sampled randomly from the entire dataset (see step 1), 10 times.

\subsubsection{Dataset II.}
Next generation sequencing promises to expand the scope of metagenomic projects by significantly increasing the number of organisms that can be sequenced from any given sample. One challenge for metagenomic analysis is the accuracy with which short reads are classified into groups representing the same or similar taxa. Improving the classification accuracy in such studies should lead to more reliable estimates of biological diversity in sequenced sample. For this reason we evaluate the ability of our model to classify reads using a real Acid Mine Drainage (AMD) metagenome (\citealp{Tyson2004}). This dataset is known to contain three dominant populations; the archaeon Ferroplasma acidarmanus and two groups of bacteria, Leptospirillum sp. groups II and III. Reads that aligned with high confidence to draft genomes of these three micro-organisms were first identified using the MUMmer algorithm (\citealp{Delcher2003}) (with the minimum length of a match set to 70\% of the full read length). A total of 20907 of these reads were found (with an average of 759 bp in length), of these 18579 aligned to Leptospirillum sp. groups II and III  genomes and 2328 to the Ferroplasma acidarmanus genome. The classification performance was evaluated at the phylum level using a training set composed of complete bacterial and archaeal genomes across 15 different phyla and 86 sequences downloaded from the NCBI RefSeq database. The 15 phyla include both the Euryarchaeota and Nitrospirae phyla to which Ferroplasma acidarmanus and Leptospirillum sp. groups II and III  belong to. The three draft reference genomes were not used as part of the training set. Selection of similarity measures prior to classification of the test examples was conducted as described in Methods section (see Selection of similarity measures prior to classification). Set A in this case consisted of complete bacterial and archaeal genomes as described above, while set B consisted of sequences of 1000 bp in length that were sampled at random from complete bacterial and archaeal genomes in the training set in such a way that for each phylum, set B consisted of genomes that were not present in set A.

\subsubsection{Dataset III.}
One of the objectives of protein sequence analysis is the inference of structure or function of unannotated protein sequences encoded in the genome. We test the ability of the CSSS model to correctly classify previously unseen protein families drawn from the Structural Classification of Proteins (SCOP) database (\citealp{Murzin1995}).  The protein dataset  consists of 4352 distinct protein sequences (ranging from 20 to 994 amino acids in length) grouped into 54 families and 23 superfamilies (\citealp{Liao2003}). The protein sequences of the 54 families were divided into test and training sets in such a way that proteins within the family are considered positive test examples while proteins outside the family but within the same superfamily are considered as a positive training examples (\citealp{Liao2003}). We note that the original dataset includes negative examples, which we did not use in our evaluation. Selection of similarity measures prior to classification of the test examples was conducted as described in Methods section. In this case the training set consisted of 1779 proteins belonging to the positive training examples which were then split into set A and set B as described in Methods section (see Selection of similarity measures prior to classification).

\end{methods}

\section{Results}
The evaluation of the classification performance on Datasets I and II was carried out using the accuracy classification score defined in eq.16 shown below, 

\begin{equation}	   
accuracy(y_i, \hat{y}_i) =  1/n_{ts} \cdot \sum\limits_{i = 0}^{n_{ts} - 1} 1( \hat{y}_i = y_i ) 
\end{equation}

where $\hat{y}_i$ is the predicted value of the $i$th sample, $y_i$ is the corresponding true value, $n_{ts}$ is the total number of test samples and 1(x) is the indicator function having a value of 1 when $\hat{y}_i$ = $y_i$ and 0 when  $\hat{y}_i$ $\neq$ $y_i$. 

As explained in the introduction section we decided to compare the results obtained in this study on Datasets I and II to six other composition and alignment based models that were developed for the classification of metagenomic data with reads (or fragments) as short as 100 bp in length. Of these six models PhymmBL (\citealp{Brady2009}) is the method closest in approach to ours since it combines scores from interpolated Markov models (IMMs) with those of BLAST resulting in a combined score that achieves higher accuracy than BLAST scores alone. 


\subsection{Dataset I  - Taxonomic classification of viral sequences}
We  evaluate the classification performance of the CSSS model (see eq.9) by predicting genera of viral DNA sequences in Dataset I (see Methods section). The training and test sets are  generated as described in the Methods section - Dataset I. The classification of test examples is then performed using the nearest neighbour algorithm (1-NN) with the combined sequence similarity scores calculated as given in eq.9. For this dataset the combined score in eq.9 is calculated based on scores obtained with the three alignment-free measures (see Methods section eqs 3, 5 and 7) and the normalized BLAST score (see eq.13). 
The value for the k-mer size is varied between 2 and 5, and the classification performance of the individual similarity measures is determined for each training and test sets as described in Dataset I, step 2 (see Methods section). The optimum value for the k-mer size is then selected based on the following two conditions {\it{i)}} best classification performance and {\it{ii)}} k-mer size has to satisfy the inequality given in eq.15. Note that the optimum k-mer sizes was estimated separately for complete viral genomes and for each of the three different viral fragment lengths (see Dataset I in Methods section).

In Table.1 we compare the classification performance of the CSSS model to five other models, PAUDA (\citealp{Huson2014}), NBC (\citealp{Rosen2011}), Kraken (\citealp{Wood2014}), PhymmBL (\citealp{Brady2009}) and RAIphy (\citealp{Nalbantoglu2011}). We note that for this dataset we could not compare the results obtained with the CSSS model to those of PhyloPythiaS (\citealp{Patil2011}) for two reasons {\it{i)}} PhyloPythiaS requires at-least 100 kb of sequence for each genus and {\it{ii)}} our training set, composed of 147 different genera, exceeds the file limit size of 10 MB imposed by the PhyloPythiaS web server. We also note that PhymmBL has been shown to perform better (see \citealp{Brady2011}) for shorter read lengths (100bp-800 bp) than both PhyloPythiaS and RAIphy. The results presented in Table.1 were obtained using identical training and test sets.

\begin{table*}[!ht]
\processtable{Shows the classification accuracy (see eq.16) for Dataset I obtained with the CSSS (1-NN classifier) and the five other models PhymmBL (\citealp{Brady2009}), NBC (\citealp{Rosen2011}), Kraken (\citealp{Wood2014}), RAIphy (\citealp{Nalbantoglu2011}) and PAUDA (\citealp{Huson2014}) when predicting 147 different viral genera across 266 viral DNA sequences as a function of the viral fragment length. \label{Tab:01}}
{\begin{tabular}{p{3.0cm}p{3.5cm}p{3.0cm}p{3.0cm}p{3.0cm}}\toprule
Classifier & & Viral fragment length & &\\
 &  Full length genomes & 1000 bp  & 500 bp  & 100 bp\\
 &  accuracy(\%) & accuracy(\%) & accuracy(\%) & accuracy(\%)\\\midrule  
CSSS  & 91.43 $\pm$ 0.99 & 70.02 $\pm$ 2.01 & 63.02 $\pm$ 1.49 & 35.94 $\pm$ 3.31 \\
PhymmBL  & 86.56 $\pm$ 2.19 & 68.90 $\pm$ 1.78 & 57.28 $\pm$ 2.09 & 29.79 $\pm$ 1.66 \\
NBC  & 74.67 $\pm$ 0.64 & 59.06 $\pm$ 1.49 & 50.39 $\pm$ 2.77 & 34.04 $\pm$ 1.53 \\
Kraken  & 48.47 $\pm$ 1.85 & 26.66 $\pm$ 1.94 &  23.07 $\pm$ 2.19 & 16.26 $\pm$ 1.40\\
RAIphy & 42.03 $\pm$ 1.56 & 30.72 $\pm$ 1.66 & 23.97 $\pm$ 1.66 & 14.06 $\pm$ 1.17\\
PAUDA  & 0.10 $\pm$ 0.15 & 6.73  $\pm$ 1.40  & 21.22  $\pm$ 1.32 & 31.89  $\pm$ 2.42\\\botrule
\end{tabular}}{}
\end{table*}

Table.1 shows that the CSSS model and PhymmBL significantly outperform other classification methods for short viral fragments (500bp - 1000bp) and complete viral genomes. Furthermore significant improvement in classification accuracy is obtained when using the CSSS model over that of PhymmBL for 100bp - 500bp viral fragments and complete viral genomes. We found no significant difference between the CSSS model and PhymmBL for 1000 bp fragments (p-value = 0.23, using the two sample t-test). Also no significant difference was found between the CSSS model and NBC (\citealp{Rosen2011}) for very short 100bp viral fragments (p-value = 0.13, using the two sample t-test). We refer the reader to the Discussion section for the explanation of these two results. Because CSSS and PhymmBL are both hybrid models that combine the alignment-based and the alignment-free/composition-based approaches, in Table.4 in Supplementary data we compare the performance of the CSSS model to that of PhymmBL when BLAST scores are used for classification alone. Both models achieve higher accuracy than BLAST scores alone (except for the CSSS model with short 100 bp fragments). From Table.4 in Supplementary data we also note that higher accuracy is achieved when classification is performed using BLAST scores alone with CSSS rather than PhymmBL, we explain the reason for this discrepancy in the Discussion section.


\subsection{Dataset II - Classification of reads from a real metagenome dataset}
For this dataset the k-mer size was set to 4 in order to satisfy the inequality in eq.15 with $L$ = 1000 bp. 
The classification performance of the CSSS model was evaluated using the training and test sets as described in Dataset II (see Methods section). The combined score in eq.9 was calculated based on scores calculated with the three alignment-free measures (see Methods section eqs 3, 5 and 7) and the normalized BLAST score (see eq.13). 
In Table.2 we compare the classification performance of the CSSS method to that of six other models on Dataset II (see Methods section). Dataset II is composed of 20907 reads (with an average of 759bp in read length) that are known to align to three genomes as described in the Methods section - Dataset II. Both CSSS and PhymmBL achieve higher level of accuracy than any other model, followed by PhyloPythiaS. PhymmBL achieves a slightly higher accuracy than CSSS for reads that align to Leptospirillum sp. groups II and III genomes (Nitrospirae phylum) while the CSSS model performs better at classifying reads that align to the Ferroplasma acidarmanus genome (Euryarchaeota phylum). Again we show in Table.5 in Supplementary data that the performance based solely on BLAST scores for the two best models (CSSS and PhymmBL) is superior for the CSSS model than PhymmBL.

\begin{table}[!ht]
\processtable{Shows the classification accuracy (see eq.16) for Dataset II obtained with the CSSS (1-NN classifier) and the six other models PhymmBL (\citealp{Brady2009}), PhyloPythiaS (\citealp{Patil2011}), NBC (\citealp{Rosen2011}), Kraken (\citealp{Wood2014}), RAIphy (\citealp{Nalbantoglu2011}) and PAUDA (\citealp{Huson2014}) when predicting the phyla for 20907 reads belonging to Leptospirillum sp. groups II and III genomes (18579 reads) and Ferroplasma acidarmanus genome (2328 reads). \label{Tab:02}}
{\begin{tabular}{p{2.6cm}p{2.6cm}p{2.6cm}}\toprule
Classifier  & Euryarchaeota accuracy(\%)  & Nitrospirae accuracy(\%)\\\midrule  
CSSS & 87.03 & 96.66\\
PhymmBL & 81.14 & 97.67\\
PhyloPythiaS & 72.76 & 95.42\\
NBC & 16.15 & 82.07\\
Kraken & 0.26 & 77.14\\
RAIphy & 1.03 & 66.99\\
PAUDA & 4.38 & 8.41\\\botrule
\end{tabular}}{}
\end{table}

\subsection{Dataset III - Classification of protein sequences}

Next we evaluated the ability of the CSSS model to classify protein sequences in Dataset III (see Methods section). Dataset III was originally created in order to evaluate methods for detecting distant sequence similarities among protein sequences as described in (\citealp{Liao2003}). The results obtained with the CSSS model are compared to those presented in \citealp{Kocsor2006} where the performance of the combined similarity measure LZW-BLAST (obtained by combining compression-based LZW and BLAST scores) was compared to that of the Smith-Waterman algorithm and two hidden Markov model-based algorithms using two types of classifiers the nearest neighbours (1-NN) algorithm and the support vector machine (SVM). Instead of calculating BLAST scores, the evaluation of the CSSS model on this protein dataset was performed using Smith-Waterman p-values, taken from \citealp{Liao2003}. The k-mer size for this dataset was set to 1 since the much larger alphabet size for protein sequences (n = 20) requires sequences of length $L$ $\geq$ 400 for the k-mer size of 2 (see eq.15) a value that is much larger then the length of many of the protein sequences in Dataset III. The combined score in eq.9 is calculated based on scores obtained using the three alignment-free measures (see Methods section eqs 3, 5 and 7) and normalized Smith-Waterman p-values (see eq.14). For the purpose of comparison with results presented in \citealp{Kocsor2006} the classification results of the CSSS model are expressed as the integral of the AUC curve shown in Figure.2 in Supplementary data (note that since Dataset III contains 54 families the maximum value for this integral is 54).


In Table.3 we show that the CSSS method achieves a slightly better performance than the Smith-Waterman p-value similarity/distance measure (using either the SVN or the 1-NN classifier) as reported in \citealp{Kocsor2006}, and performs much better than the combined LZW-BLAST similarity measure with the 1-NN classifier also reported in \citealp{Kocsor2006}.

\begin{table}[!ht]
\processtable{Shows the classification performance on protein domain sequences for the CSSS model (1-NN classifier) with the k-mer size = 1 (see Results section), expressed as the integral of the AUC curve shown in Figure.2 in Supplementary data, and similarity/distance measures presented in \cite{Kocsor2006} (marked with an *). Since Dataset III contains 54 protein families the maximum value for the integral of the AUC curve is 54 which correspond to all 54 protein families being classified without error.\label{Tab:03}}
{\begin{tabular}{p{4.0cm}p{2.6cm}p{1.1cm}}\toprule
Similarity/Distance measure & Classification Method & \\
 & SVM &  1-NN\\\midrule
Smith–Waterman p-value* & 48.66 & 50.22\\
LZW-BLAST* & 49.0 & 37.18\\
CSSS & n.a & 50.64\\\botrule
\end{tabular}}{}
\end{table}
\section{Discussion}
Sequence comparison is at the core of many bioinformatics applications such as metagenomic classification, protein sequence and function characterization, and phylogenetic studies to name a few. In many of these applications the alignment-based sequence comparison is widely used, but this does not come without some limitations. One important limitation is that the alignment-based similarity measure might give erroneous information when used with sequences that have undergone some type of sequence rearrangement. Alignment-free similarity measures offer an alternative to the alignment-based ones in that they are unaffected by such genetic processes. 

In this study we propose a model that combines similarity scores obtained with alignment-based and alignment-free sequence similarity measures (see eq.9) to gain additional discriminatory information about sequences and to improve their characterization. 

In Tables 1 and 2 we show that our approach performs better than most of the other methods used in this study when predicting genera of unknown viral sequences (i.e. sequences that are not part of the training set as described in Dataset I) or when predicting phyla of metagenomic sequences. The main conceptual difference between the CSSS model and the other classification methods used in this study, at the exception of PhymmBL, is that the CSSS model combines similarity scores obtained with both the alignment-based and the alignment-free sequence similarity measures while the other models rely on either one of these two approaches. Thus NBC (\citealp{Rosen2011}), RAIphy (\citealp{Nalbantoglu2011}) and PhyloPythiaS (\citealp{Patil2011}) rely on the alignment-free composition-based approaches (using k-mer frequencies or k-mer counts), PAUDA (\citealp{Huson2014}) relies on the alignment-based approach and Kraken (\citealp{Wood2014}) on the exact alignment of k-mers. Although in some respects our approach is similar to that of PhymmBL, since both methods combine scores calculated using different types of similarity measures (PhymmBL uses BLAST scores and Interpolated Markov Models scores (IMM) (\citealp{Salzberg1998})), there are two main differences that can explain the results obtained with Dataset I shown in Table.1.

First the CSSS model uses four different similarity measures, so that if sufficiently independent one from another, their combined additive effect could confer a greater discriminatory power than the two similarity scores combined by PhymmBL. In Table.6 in Supplementary data we show the classification accuracy of individual similarity measures used by CSSS and PhymmBL models as a function of the viral fragment length.

While the classification performance of the ED (see eq.3) and JSD (see eq.5) measures are very similar, the classification performance of the CB (see eq.7) measure drops rapidly below 10\% as the length of viral fragments decreases. If however we perform the classification on full length viral genomes (see Dataset I in Methods section) we find that the CB measure improves the performance by as much as 5.79\% when combined with the other three measures (ED, JSD and BLAST). This shows that the CB measure contains significant additional information, only for sequences that are similar in length to those in the training set, that is complementary to the information contained by the other three measures. This drop in performance of the CB measure as a function of the fragment length, relative to the length of the genomes in the training set, explains also the smaller difference in performances observed between the CSSS and PhymmBL models when classifying longueur reads in Dataset II (see Methods section) shown in Table.2.

Since the ED and JSD measures show similar classification performances we investigated the degree of independence of these two measures by performing a principal components analysis (PCA) of the similarity scores obtained using viral genomes from Dataset I. We found that the first component (i.e. PCA1) is strongly associated with the ED measure in test samples while the second component (i.e. PCA2) is strongly associated with the JSD measure, a result that is independent of the viral fragment length as shown in Figure.3 in Supplementary data. These results indicate that these two measures can be considered as orthogonal and thus not correlated, with the ED measure accounting for most of the variation across viral genomes in test samples. To further determine the effect of these two measures on the classification performance we removed each measure from the model one at the time and then recalculated the accuracy scores. We found that for full viral genomes the effect of removing the ED measure reduced the classification performance significantly by 5\% while removing the JSD measure reduced it only slightly (0.25\%). However in the case of shorter viral fragments dropping either one of these two measures from the model did not produce any significant change to the performance, while removing both produced a significant drop in performance (up to 3\% for 1000 bp reads). In the light of these results we conclude that both of these measure contain complementary information that is useful for characterizing viral sequences.


The second important difference between our model and PhymmBL is in the weighting scheme used. In the PhymmBL model the weights assigned to each similarity measure (i.e. Combined score = IMM + 1.2(4 - log(E)), where IMM is the score of the best matching IMM and E the smallest E-value returned by BLAST) have the same value for all test examples, in the CSSS model weights are determined independently for each test example based on the discriminatory ability of each measure using the training set (see eq.9). Having different sets of weights for different test samples (i.e. test sequences) should improve the classification performance since some sequences will be better characterized with one type of similarity measure than another. 
Another important difference between these two methods is in the classification performance using BLAST results alone. As shown in Tables 4, 5 and 6 in Supplementary data we found that a significant improvement in classification is obtained when the BLASTN algorithm is used instead of mega-BLAST, the algorithm used by PhymmBL. BLASTN is more sensitive than mega-BLAST because it uses a shorter word size (default value of 11) that makes it better at finding related nucleotide sequences between more divergent biologically sequences since the initial exact match can be shorter. 

We found that for very short viral fragments (100bp in length) the CSSS model performs better than PhymmBL and achieves slightly better accuracy (but not significant p-value = 0.13, using the two sample t-test) than the NBC model, as shown in Table.1. By examining the individual performance of the sequence similarity measures used by the CSSS model we found that the composition and compression based similarity measures are more affected by the shorter fragment size than the alignment based one, as shown in Table.6 in Supplementary data. Despite this drop in performance (of the composition and the compression based similarity measures) for short 100bp viral fragments, by virtue of combining different similarity measures the CSSS model still achieves better performance than the alignment based method PAUDA (p-value = 0.008, using the two sample t-test) or the hybrid PhymmBL (Phymm + BLAST) (p-value = 0.0001, using the two sample t-test) and performs equally well as the best composition based model used in this study, namely NBC.

In the Results section we have shown that our approach can also be used effectively for protein sequence classification. In Table.3 we show that our model outperforms a similar but simpler LZW-BLAST 1-NN model \citep{Kocsor2006}. The main differences between these two approaches are the number of similarity measures used (frequency-based measures such as those given in eqs.3 and 5 were not used in \citealp{Kocsor2006}), a different method with which similarity measures are combined and Smith-Waterman scores (p-values) instead of BLAST scores. Without using a weighting scheme the LZW-BLAST method uses a simple multiplication rule to combine the LZW and BLAST scores (\citealp{Kocsor2006}). We found that the multiplication rule used in \citealp{Kocsor2006} performs significantly better in combination with an SVM rather than a nearest neighbour classifier. The model proposed in this study performs better than the SVM (LZW-BLAST) model reported in \citealp{Kocsor2006} and slightly better than the 1-NN (Smith-Waterman p-value) as shown in Table.3. We attribute this smaller gain in classification performance to the short protein sequences in Dataset III which pose a greater challenge to the three alignment-free similarity measures examined in this study. 

As shown in eq.9 our model combines similarity scores using a linear combination of vectors (equivalent to calculating a weighed arithmetic mean of scores obtained with each individual similarity measure). We did explore combining similarity scores using a different multiplicative model which we found to significantly under-perform (in combination with the nearest neighbour classifier) when used on datasets presented in this study.

Finally our approach can be easily extended to any number of additional similarity measures (such as the IMMs used by PhymmBL) that might produce additional gain in discriminatory information about sequences and thus improve the overall classification performance. Therefore, future work will include assessing the performance of additional similarity measures that could be integrated into our model.

\section*{Acknowledgement}
This work was conducted with the support of the Ontario Institute for Cancer Research through funding provided by the government of Ontario to the authors.



\bibliographystyle{bioinformatics}
\bibliography{csss}

%

\end{document}